\documentclass[%
 amsmath,amssymb,
 aps,
]{revtex4-1}

\usepackage{graphicx}% Include figure files
\usepackage{dcolumn}% Align table columns on decimal point
\usepackage{bm}% bold math

\begin{document}

\title{The topological order in loop quantum gravity \footnote{Essay written for the Gravity Research Foundation 2023 Awards for Essays on Gravitation.}}
\author{Jingbo Wang}
\email{shuijing@mail.bnu.edu.cn}
\affiliation{Department of Physics and Electronic Engineering, Hebei Normal University for Nationalities, Chengde, 067000, China}
 \date{\today}
\begin{abstract}
Topological order (long-range entanglement) is a new type of order that beyond Landau's symmetry breaking theory. This concept plays important roles in modern condensed matter physics. The topological entanglement entropy provides a universal quantum number to characterize the topological order in a system. The topological entanglement entropy of the BTZ black hole was calculated and found that it coincides with that for fractional quantum Hall states. So the BTZ black hole has the same topological order with the fractional quantum Hall state. We show that the four dimensional black hole can also have topological order, by showing that the topological entanglement entropy of black holes is non-zero in loop quantum gravity. We also consider the Hamiltonian constraint in loop quantum gravity. By comparing with the BF theory, we found that the physical Hilbert space for loop quantum gravity is degenerate on non-trivial manifold, which indicate the existence of topological order in loop quantum gravity. We advise to bring in the methods and results in string-net condensation to loop quantum gravity to solve some difficult problems.
\end{abstract}

\pacs{04.70.Dy,04.60.Pp}
 \keywords{Loop quantum gravity; topological order; topological entanglement entropy; Chern-Simons theory}
\bibliographystyle{unsrt}
\maketitle

\section{Introduction}
Topological order \cite{to1} is a new kind of order that beyond Landau's symmetry breaking theory description. Instead it can be described by topological quantum field theory. Topological order can be probed/defined by three essential properties \cite{to2}:
\begin{enumerate}
  \item  the ground state degeneracy on a torus (or other spaces with non-trivial topology);
  \item  the non-Abelian geometrical phase of those degenerate ground states (which form a representation of the modular group $SL(2,Z)$ on a torus);
  \item  the gapless edge modes.
\end{enumerate}
Those properties are all robust against any small perturbations. Microscopically topological order is just pattern of long range entanglement. Topological order can produce quasi-particles with fractional quantum numbers and fractional/Fermi statistics, robust gapless boundary modes and emergent gauge excitations. So they can provide an unification of gauge interaction and Fermi statistics \cite{wen3}. A typical system which has topological order is the fractional quantum Hall states.

It was shown that quantum gravity in $AdS_3$ spacetime have topological orders \cite{wangto1}. Actually it was shown that the quantum gravity in $AdS_3$ spacetime satisfy all the above three properties. For higher dimensional black holes, such as Kerr black holes in four dimension, the boundary modes can be described by massless scalar field, which is also the same as higher dimensional topological insulators \cite{wangbms4}. So it was conjectured that black holes in higher dimensions can also have topological orders. In the next section, we will show that the four-dimensional black holes indeed have topological order. It is shown that the black hole has a non-zero topological entanglement entropy, which indicate the existence of the topological order.

\section{The topological orders for 4D black holes in loop quantum gravity}
Loop quantum gravity is a main candidate for quantum gravity. It can eliminate the bing bang singularity in some simple cosmological models, and also give the microscopic explanation for the black hole entropy. In loop quantum gravity, the degrees of freedom on the horizon of black holes can be described by Chern-Simons theory with $U(1)$ \cite{abck1,ash1} or $SU(2)$ \cite{kaul1,kaul2,enpp1,enpp2} gauge group. Also one can use the $SO(1,1)$ BF theory to describe those degrees of freedom \cite{wmz1,wh1,ma1}. In this paper, we will just consider the Chern-Simons theory. By counting the admirable boundary states one can get the Bekenstein-Hawking area entropy for black holes \cite{bhcount1}, for a detail counting for different theories, see Ref.\cite{bhcount2}. The logarithmic correction is different for those two groups, which is $-\frac{1}{2}\log S_{BH}$ for the $U(1)$ group and $-\frac{3}{2}\log S_{BH}$ for the $SU(2)$ group. This difference can be due to the gauge-fixing from the $SU(2)$ to $U(1)$ group \cite{bkm1}. Those two terms were obtained through complicated mathematical tools. In this section, we will show that the logarithmic corrections have simple explanation: they are just the topological entanglement entropy for the Chern-Simons theory with different gauge group.

A universal quantum number to characterize the topological order in a system can be provided by the topological entanglement entropy (TEE)\cite{tee1,tee2}. For a topological ordered system, the entanglement entropy satisfies the area law
\begin{equation}\label{1}
  S\sim \alpha A-\gamma,
\end{equation}
where $A$ is the area of the surface, and $\alpha$ is a parameter that depends on the system, $\gamma>0$ is a UV-insensitive term. The first term shows the short-range entanglement near the surface, and the non-zero second term indicates the existence of long-range entanglement that origins the topological order. This term is called the topological entanglement entropy. For example, the TEE was calculated for Euclidean $AdS_3$ spacetime using surgery method\cite{sun1}. It is related to the so-called ``total quantum dimension" $\mathcal{D}$ of the system with $\gamma=\ln \mathcal{D}$. The total quantum dimension measure the number of the quasi-particles and is defined by
\begin{equation}\label{2}
  \mathcal{D}=\sqrt{\sum_a d_a^2},
\end{equation}
where $d_a$ is the quantum dimension for the quasi-particle type $a$.

For $U(1)$ Chern-Simons theory description of the black hole with level $k_1=\frac{a_H}{4\pi l_p^2 \beta}$, the quantum dimension for each quasi-particle is $d_a=1$ for $a=1,2,\cdots,k$, so the TEE for this theory is
\begin{equation}\label{3}
  \gamma_1=\ln \mathcal{D}=\ln \sqrt{\sum_a d_a^2}=\ln \sqrt{k}=\frac{1}{2}\ln S_{BH}-\frac{1}{2}\ln (\pi \beta),
\end{equation}
which gives the right logarithmic correction term $-\frac{1}{2}\log S_{BH}$.

For $SU(2)$ Chern-Simons theory with level $k_2=\frac{a_H}{4\pi l_p^2 \beta(1-\beta^2)}$, there are $k+1$ types of quasi-particles labeled by $j=0,1,\cdots,k$ (for a more detail, see for example Ref.\cite{gsd1}). The quantum dimension for each type is given by $d_j=\frac{\sin{\frac{(j+1)\pi}{k+2}}}{\sin{\frac{\pi}{k+2}}}$, so the TEE for this theory is
\begin{equation}\label{4}
  \gamma_2=\ln \mathcal{D}=\ln \sqrt{\sum_j d_j^2}=\ln \sqrt{\frac{k_2+2}{2\sin^2{\frac{\pi}{k_2+2}}}}\approx\frac{1}{2}\ln \frac{(k_2+2)^3}{2\pi^2}\approx \frac{3}{2}\ln S_{BH}-\frac{1}{2}\ln (2\pi^5 \beta^3 (1-\beta^2)^3),
\end{equation}
for very large $k$. It also matches the result from the states-counting method $-\frac{3}{2}\log S_{BH}$.

In loop quantum gravity, the black hole entropy is essentially the entanglement entropy of gravitational field through the horizon \cite{bhee1,bhee2,bhee3}. In this section, we show that the logarithmic correction terms can be identified with the topological entanglement entropy, which indicate that the black hole have the topological order. The $SU_k(2)$ Chern-Simons theory with large $k$ can be used for universal quantum computing, so it suggests that the black hole can be considered not just as a quantum computer but also as a topological quantum computer.

In Ref.\cite{lqgqc1} the non-abelian anyons on the horizon of black hole in loop quantum gravity are investigated, and suggest a connection to the topological quantum computation. The conjectured relation between large diffeomorphisms transformation and the log-correction term \cite{btzlqg1} can be also confirmed. The log-correction is identified with the topological entanglement entropy, which indicate the topological order. According to the property (2) of topological order, the degenerate ground states form the representation of the mapping class group, which is formed by the large diffeomorphisms transformation. So the large diffeomorphisms transformation act as symmetry (not gauge) transformation on the physical states.
\section{The topological order in loop quantum gravity}
In loop quantum gravity, the general relativity is rewritten as a background-independent $SU(2)$ Yang-Mills gauge theory. The canonical variables are $(A^i_a,E^a_i)$, where $A^i_a$ are Ashtekar-Barbero connections, and $E^a_i$ are triads with density weight one. The theory subjects to three constraints:
\begin{align}\label{5}\begin{split}
  & G_i=\frac{1}{8\pi \beta G}\mathcal{D}_a E_i^a:=\frac{1}{8\pi \beta G}(\partial_a E_i^a+\epsilon_{ijk}A_a^j E^{ak})=0,\\
  & C_a=\frac{1}{8\pi \beta G}E_i^b F^i_{ab}=0,\\
  & C= \frac{1}{16\pi G} \frac{E_i^a E_j^b}{\sqrt{E}} (\epsilon^{ij}_{\ \ k}F^k_{ab}-(1+\beta^2)(K^i_a K^j_b-K^j_a K^i_b))=0,
\end{split}\end{align}
which are Gauss constraint, diffeomorphim constraint and Hamiltonian constraint respectively. The Gauss constraint can be solved by so-called ``spin network states" which provide the basis for the kinematical Hilbert space. The diffeomorphim constraint can also be solved by ``the group averaging procedure''. The Hamiltonian constraint contains two terms, and the first one is called the Euclidean term $C_{Eucl}=\frac{1}{16\pi G} \frac{E_i^a E_j^b}{\sqrt{E}} \epsilon^{ij}_{\ \ k}F^k_{ab}$, since this is the only term for Euclidean general relativity. The Hamiltonian constraint is difficult to solve due to its complicate form. Actually it is one central problem in loop quantum gravity.

To handle this difficult problem, it is useful to consider a similar but simpler theory--the $SU(2)$ BF theory \cite{tqft1,bf1}, which is a topological quantum field theory. The action for the BF theory is
\begin{equation}\label{6}
  S[B,A]=\int tr(B \wedge F[A]).
\end{equation}
The phase space is spanned by canonical variables $(A^i_a, E_i^a=\frac{1}{2}\epsilon^{abc} B_{ibc})$. This theory has two types of constraints, the Gauss constraint and the ``Hamiltonian constraint",
\begin{align}\label{7}\begin{split}
  & G_i=\mathcal{D}_a E_i^a=0,\\
  & C^a_i=\frac{1}{2}\epsilon^{abc} F_{ibc}=0.
 \end{split}\end{align}
The second constraint is also called the ``flat constraint" since it mean that the connection should be flat (its curvature is zero). It is easy to see that if one solve the flat constraint, the diffeomorphim constraint and Hamiltonian constraint in loop quantum gravity are solved simultaneous for Euclidean general relativity. That is, the physical Hilbert space for BF theory is a subspace of physical Hilbert space for Euclidean loop quantum gravity.

Actually the physical Hilbert space for BF theory is well understand. A lattice version for the BF theory is given by the string-net model (Levin-Wen model \cite{lw1} and Walk-Wang model \cite{ww1}). The Hamiltonian for the string-nets is exactly soluble. It takes the form
 \begin{equation}\label{8}
   H=-\sum_I Q_I-\sum_p B_p, \quad B_p=\sum_{s=0}^N a_s B_p^s,\quad a_s=\frac{d_s}{\sum_{i=0}^N d_i^2},
\end{equation}
where the sums rum over vertices $I$ and the plaquettes $p$ of the lattice. The first term $Q_I$ is an electric charge operator, and the second term $B_p$ is a linear combination of magnetic flux operators $B_p^s$. They are correspond to the Gauss constraint and flat constraint in BF theory respectively. The string-net model can have topological orders \cite{gww1}, that is, the ground states are degenerate on the non-trivial manifolds. So we can derive that the Euclidean loop quantum gravity can also have topological orders. For Lorentzian loop quantum gravity, if we assume that the second term in Hamiltonian constraint is a perturbative term, and doesn't change the universal class of the theory, then the Lorentzian loop quantum gravity can also have topological orders.

In topological phase of matter, a widely observed phenomenon is the boundary-bulk correspondence \cite{bounbulk1,bounbulk2}, which relates the topological structure of the bulk states to the presence of protected zero-energy boundary states. This boundary-bulk correspondence can be considered as a weak version of the holographic principle \cite{hp1,hp2} in gravity theory. Actually in Ref.\cite{hp1}, it was suggest that ``quantum gravity should be described entirely by a topological quantum field theory". And topological quantum field theory can describe the low energy behavior of topological phase of matter.
On the other hand, the studies in loop quantum gravity can also be applied to the researches in string-net condensation theory. A main problem in the unification through string-net condensation is how to construct the graviton \cite{gw1}. But in loop quantum gravity, there are some attempts to construct graviton form loop or spin network \cite{gw2,gw3,gw4}, which may be useful for string-net condensation.

In summary, we can use the methods in string-net condensation to study the spin network states in loop quantum gravity. If space has topological order, it can support gauge field and Fermion field from simple bosonic qubits. It may solve the origin of elementary particles in the standard model, and resents an unification of matter and information \cite{wen3}.

\section{Conclusion}
In this paper, we investigate the topological order in loop quantum gravity. Firstly we show that the black hole have topological order by showing that the topological entanglement entropy for black holes are non-zero. Secondly we study the topological order of general states in loop quantum gravity. By showing that the physical Hilbert space of BF theory is subspace of physical Hilbert space of Euclidean loop quantum gravity, and are degenerate on non-trivial manifold, we prove that the Euclidean loop quantum gravity also have topological order, just like the BF theory.

Topological order is an important new concept in modern condensed matter physics. Through the study of topological order in loop quantum gravity, we can bring in the methods and ideas in condensed matter physics to the realm of loop quantum gravity, and solve some difficult problems in the loop quantum gravity.

\acknowledgments
 This work is supported by Doctoral Startup Fund of Hebei Normal University for Nationalities and Hebei Provincial Department of Education Science and Technology Research Program with Grant No. QN2023144.

%\bibliography{essay2021}

\end{document}